\documentstyle[12pt]{article}

\newcommand{\be}{\begin{equation}}
\newcommand{\ee}{\end{equation}}
\newcommand{\ba}{\begin{eqnarray}}
\newcommand{\ea}{\end{eqnarray}}
\newcommand{\x}{\vec{x}}
\newcommand{\k}{\vec{k}}
\newcommand{\vP}{\vec{\Phi}}
\newcommand{\vp}{\vec{\pi}}

\newcommand{\h}{\hbar}
\begin{document}
\begin{flushright}
{PITT-94-01}\\
{LPTHE-94-01}\\
{CMU-HEP-94-05}\\
{DOR-ER/40682-59}\\
{1-19-94}
\end{flushright}
\begin{center}
{\bf CAN DISORDERED CHIRAL CONDENSATES FORM? A DYNAMICAL PERSPECTIVE}
\end{center}
\begin{center}
{{\bf D. Boyanovsky$^{(a)}$, H.J. de Vega$^{(b)}$ and R. Holman$^{(c)}$}}
\end{center}
\begin{center}
{\it (a)  Department of Physics and Astronomy, University of
Pittsburgh, Pittsburgh, PA. 15260, U.S.A.} \\
{\it (b)  Laboratoire de Physique Th\'eorique et Hautes Energies$^{[*]}$
Universit\'e Pierre et Marie Curie (Paris VI),
Tour 16, 1er. \'etage, 4, Place Jussieu
75252 Paris, Cedex 05, France}\\
{\it (c) Department of Physics, Carnegie Mellon University, Pittsburgh,
PA. 15213, U. S. A.}
\end{center}
\begin{abstract}
We address the issue of whether a region of disordered chiral condensate
 (DCC),
in which the chiral condensate has components along the pion directions,
 can form.
We consider a system going through the chiral phase transition either via
a quench, or via relaxation of the high temperature phase to the low
temperature one within a given time scale (of order $\sim 1 \rm{fm/c}$).
 We use
a density matrix based formalism that takes both
thermal and quantum fluctuations into account non-perturbatively
 to argue that if the $O(4)$ linear  sigma
model is the correct way to model the situation in QCD, then it is very
unlikely at least in the Hartree approximation,
 that a large ($> 10\ \rm{fm}$) DCC region will form. Typical
 sizes of
such regions are $\sim 1 -2 \ \rm{fm}$ and  the density of pions in such
 regions is
at most of order $\sim 0.2 / \rm{fm}^3$.
We end with some speculations on how
large DCC regions may be formed.

\end{abstract}

\newpage

\section{\bf Introduction}

The proposition has been put forth recently that regions of misaligned
chiral
vacuum, or disordered chiral condensates (DCC) might form in either
ultra-high
energy or heavy nuclei collisions\cite{dcc1,dcc2,dcc3,dcc4}.
 If so, then this would be
 a
striking probe of the QCD phase transition. It might also help
 explain\cite{dcc5,dcc6}
 the
so-called Centauro and anti-Centauro events observed in high-energy cosmic
ray
experiments\cite{dcc7}.

How can we tell, from a theoretical standpoint, whether or not we should
expect
a DCC to form? Clearly, investigating QCD directly is out of the question
 for
now; the technology required to compute the evolution of the relevant
 order
parameters directly from QCD is still lacking, though we can use lattice
calculations for hints about some aspects of the QCD phase transition.
 What
we
need then is a model that encodes the relevant aspects of QCD in a
 faithful
manner, yet is easier to calculate with than QCD itself.

Wilczek and Rajagopal\cite{wilraj1} have argued
 that the $O(4)$ linear $\sigma$-model is such a model.
 It lies within the same
 {\em
static} universality class as QCD with {\em two} massless quarks,
 which is a
fair approximation to the world at temperatures and energies below
$\Lambda_{QCD}$. Thus work done on the $n=4$ Heisenberg ferromagnet can
 be
used
to understand various static quantities arising at the chiral phase
transition.

One conclusion from reference(\cite{wilraj1}) was that, if as the critical
temperature for the chiral phase transition was approached from above the
system remained in thermal equilibrium, then it was very unlikely that a
large
DCC region, with its
concomitant
biased pion emission, would form.

 The point
was that the correlation length $\xi=m_{\pi}^{-1}$
 did {\em not} get large
compared
to the $T_c^{-1}$. A more quantitative criterion involving the
 comparison of
the energy in a correlation volume just below $T_c$ with the $T = 0$
 pion mass
(so as to find the number of pions in a correlation volume) supports the
conclusion that as long as the system can equilibrate, no large regions of
DCC will form.

The only option left, if we want to form a DCC, is to insure that the
 system
is
far out of equilibrium. This can be achieved by {\em quenching} the
 system
(although Gavin and M\"{u}ller\cite{gavin} claim that {\em annealing}
 might also work).
What this means is the following.
 Start with the system in equilibrium at a
temperature above $T_c$. Then suddenly drop the temperature to zero. If the
rate at which the temperature drops is much faster than the rate at which
the
system can adapt to this change, then the state of the system after the
quench
is such that it is still in the thermal state at the initial temperature.
However, the {\em dynamics} governing the evolution of that initial state is
now driven by the $T=0$ Hamiltonian. The system will then have to relax
from
the initial state, which is {\em not} the ground state of the Hamiltonian to
the zero temperature ground state. During this time, it is expected that
regions in which the order parameter is correlated will grow. We can then
hope
that the correlation regions will grow to be large enough to contain a
 large
number of pions inside them.

The possibility that the chiral phase transition might occur following
a  quench in heavy-ion collisions was explored by Wilczek and Rajagopal
 in
reference(\cite{wilraj2}).
 They argue there that long wavelength fluctuations in
the
pion fields can develop after the quench occurs. Modes with wavenumbers
 $k$
smaller than some critical wavenumber $k_{\rm{crit}}$ will be unstable
 and
regions in which the pion field is correlated will grow in spatial
 extent for
a period of time. The essence of this mechanism is that the pions are the
would-be Goldstone particles of spontaneous chiral symmetry breaking. In
the
absence of quark masses, the pions would be exactly massless when the
$\vec{\pi}$ and $\sigma$ fields are in their ground state. However, during
the
quench, the $\sigma$ field is displaced from its zero temperature minimum so
that the required cancellation between the negative bare $\rm{mass}^2$ term
in the Lagrangian and the $\rm{mass}^2$ induced through the pion
 interactions with
the $\sigma$ condensate does not occur. This then allows some of the pion
momentum modes to propagate as if they had a {\em negative} $\rm{mass}^2$,
thus
causing exponential growth in these modes.
In Ref.(\cite{wilraj2}), the {\em classical} sigma model was simulated.
The correlation
functions were taken as spatial averages and
the
expectations for the growth of various pion field momentum modes were borne
out.
More recently  however, Gavin, Gocksch and Pisarski\cite{pisarski}
 have concluded that
the strongly coupled linear sigma models {\it does not} produce large
correlated domains of pions. These authors also performed a numerical
simulation of the {\it classical} equations of motion.
 However, there are already hints from previous
 work\cite{boysinlee,danhec}
that both
{\em quantum and thermal} effects, may be important in
determining the growth of correlation regions in a field theory. It is to
this
calculation that we devote the rest of this paper. In the next section, we
will
develop the formalism necessary to take the quantum and thermal
 fluctuations of the
fields
into account. Having done that, we turn to the actual numerical solution
 of the
equations we find, and use these solutions to calculate the equal time
two-point correlation function for the pions.
 It will be clear after doing
this  that in the case in which the system is either quenched or
relaxed from a high temperature phase
whose temperature is larger than the critical temperature,
the correlation regions are never large enough for the
 correlations in
pions to be observed. We strengthen this conclusion by computing
 the number of
pions in the correlation volume. We then end with some speculations
concerning
possible ways in which a large DCC region might form.

\section{\bf The $O(4)$ $\sigma$-Model Out of Equilibrium}

Our strategy is as follows. We will use the techniques developed by us
previously\cite{frw} and use the functional
Schr\"{o}edinger representation,
in which the time evolution of the system is represented by the
 time evolution
of its {\em density matrix}.

The next step is to evolve the density matrix in time from this initial
 state
via the quantum Liouville equation:

\be
i\h \frac{\partial \rho (t)}{\partial t} = [H, \rho(t)],
\ee
where $H$ is the Hamiltonian of the system {\em after} the quench. Using
 this
density matrix, we can, at least in principle, evaluate the equal time
correlation function for the pion fields, and observe its growth with
 time.

Let us now implement this procedure. We start with the sigma model
 Lagrangian
density

\ba
{\cal L} & = & \frac{1}{2}\partial_{\mu}\vec{\Phi}\cdot
\partial^{\mu}\vec{\Phi}
- V(\sigma, \vp) \\
V(\sigma, \vp)
         & = & \frac{1}{2} m^2(t) \vP \cdot \vP + \lambda
(\vP \cdot \vP)^2 - h \sigma \label{potential}
\ea
where $\vP$ is an $O(N+1)$ vector,
$\vP= (\sigma, \vp)$ and $\vp$ represents
the $N$ pions.

The linear sigma model is a low energy effective theory for an
$SU_{\rm{L}}(2)\times SU_{\rm{R}}(2)$
(up and down quarks) strongly interacting
theory. It may be obtained as a Landau-Ginzburg effective theory from a
Nambu-Jona-Lasinio model\cite{klebansky}.
In fact, Bedaque and Das\cite{bedaque}
have studied a quench starting from an $SU_{\rm{L}}(2)\times SU_{\rm{R}}(2)$
Nambu-Jona-Lasinio model.

We have parametrized the dynamics of the cooling down process in terms of a
time dependent mass term. We can use this to describe the phenomenology
 of
either a sudden quench where the mass$^2$ changes sign
instanteneously or that of a relaxational
process in which the mass$^2$ changes sign on a time scale
 determined by the
dynamics. In a heavy ion collision, we expect this relaxation time scale
 to be
of the order of $\tau \sim 0.5-1 \rm{fm/c}$.
The term $h \sigma$ accounts for the explicit breaking of chiral symmetry  due
to the (small) quark masses. We leave $N$ arbitrary for now, though at
 the end we will take $N=3$.

 Our first order of business is to identify the correct order
parameter for the phase transition and then to obtain its equation of
 motion.
Let us define the fluctuation field operator $ \chi(\x,t)$ as

\be
\sigma = \phi(t) + \chi(\x,t),
\ee
with $\phi(t)$ a c-number field defined by:

\ba
\phi(t) & = & \frac{1}{\Omega} \int d^3 x \langle \sigma(\x) \rangle
 \nonumber \\
        & = & \frac{1}{\Omega} \int d^3 x \frac{{\rm{Tr}}(\rho(t)
\sigma(\x))}{{\rm{Tr}}\rho(t)} .
\ea
Here $\Omega$ is the spatial volume we enclose the system in. The fluctuation
field $\chi(\x,t)$ is defined so that (i) $\langle \chi(\x,t) \rangle=0$,
and
(ii) $\dot{\chi}(\x,t)=-\dot{\phi}(t)$. Making use of the Liouville equation
for
the density matrix, we arrive at the following equations:

\ba
\dot{\phi}(t) & = & p(t) = \frac{1}{\Omega} \int\ d^3 x\
\langle \Pi_{\sigma}(\x) \rangle \label{momexp} \\
\dot{p}(t)    & = & -\frac{1}{\Omega}\int d^3x\
\langle\frac{\delta V(\sigma, \vp)}{\delta \sigma(\vec{x})} \rangle ,
\ea
where $\Pi_{\sigma}(\x)$ is the canonical momentum conjugate to
 $\sigma(\x)$.

The derivative of the potential in the equation for $\dot{\pi}(t)$ is to
be evaluated at $\sigma = \phi(t) + \chi(\x,t)$. These equations can be
combined into a single one describing the evolution of the order parameter
$\phi(t)$:

\be
\ddot{\phi}(t) + \frac{1}{\Omega}\int d^3x\
\langle
\frac{\delta V(\sigma, \vp)}{\delta \sigma(\vec{x})}
\rangle
\mid_{\sigma = \phi(t) +
\chi(\x,t)} \rangle = 0.
\ee

To proceed further we have to determine the density matrix. Since the
Liouville equation is first order in time we
need only specify $\rho(t=0)$.
At this stage
we could proceed to a perturbative description of the
dynamics (in a loop expansion).

However, as we learned previously in a
similar situation\cite{boysinlee,danhec},
the non-equilibrium
dynamics of the phase transition cannot be studied within perturbation
theory.

Furthermore,
since the quartic coupling of the linear sigma model $\lambda$
must be large ($\lambda \approx
4-5$ so as to reproduce the value of $f_{\pi}
\approx 95$ Mev with a ``sigma mass'' $\approx$ 600 Mev), the linear
sigma model is a {\it strongly} coupled theory and any type
 of perturbative
expansion will clearly be unreliable.
Thus,
following our previous work\cite{boysinlee,frw} and
the work of Rajagopal and Wilczek\cite{wilraj2} and Pisarski\cite{pisarski}
 we invoke a Hartree approximation.

In the presence of a vacuum expectation value, the Hartree factorization is
somewhat subtle. We will make a series of {\it assumptions} that we feel are
quite reasonable but which, of course, may fail to hold under some
circumstances and for which we do not have an {\it a priori} justification.
These are the following: i) no cross correlations between the pions and the
sigma field, and ii) that the two point correlation functions of the pions are
diagonal in isospin space, where by isospin we now refer to the unbroken
$O(N)\ (N=3)$
symmetry under which the pions transform as a triplet. These assumptions
lead to the following Hartree factorization of the non-linear terms in the
Hamiltonian:

\ba
\chi^4 & \rightarrow & 6 \langle \chi^2 \rangle +\mbox{constant}
\label{hartreechi4} \\
\chi^3 & \rightarrow & 3 \langle \chi^2 \rangle \chi
\label{hartreechi3} \\
(\vec{\pi}\cdot\vec{\pi})^2
       & \rightarrow & (2+\frac{4}{N})\langle \vec{\pi}^2 \rangle
\vec{\pi}^2 + \mbox{constant} \label{hartreepi4} \\
\vec{\pi}^2 \chi^2
       & \rightarrow & \vec{\pi}^2 \langle \chi^2 \rangle+ \langle
\vec{\pi}^2 \rangle \chi^2 \label{hartreepi2chi2} \\
\vec{\pi}^2 \chi
       & \rightarrow & \langle \vec{\pi}^2 \rangle \chi ,
\label{hartreepi2chi}
\ea
where by ``constant'' we mean the operator independent expectation
values of the composite operators. Although these will be present as
operator independent terms in the Hamiltonian, they are c-number terms
and will not enter in the time evolution of the density matrix.

It can be checked that when $\phi =0$ one obtains the
$O(N+1)$ invariant Hartree factorization.

In this approximation the resulting Hamiltonian is quadratic, with a
 linear term  in $\chi$:

\be
H_H(t) = \int d^3x \left\{ \frac{\Pi^2_{\chi}}{2}+
\frac{\vec{\Pi}^2_{\pi}}{2}+\frac{(\nabla \chi)^2}{2}+
\frac{(\nabla \vec{\pi})^2}{2}+\chi {\cal{V}}^1(t)+
\frac{{\cal{M}}^2_{\chi}(t)}{2}
\chi^2+\frac{{\cal{M}}^2_{\pi}(t)}{2}\vec{\pi}^2 \right\} .
 \label{hartreeham}
\ee

Here $\Pi_{\chi}, \ \vec{\Pi}_{\pi}$ are the canonical momenta
conjugate to $\chi(\x), \ \vec{\pi}(\x)$ respectively and
${\cal{V}}^1$ is recognized as the derivative of the Hartree
``effective potential''\cite{moshe,camelia} with respect to $\phi$
(it is the derivative of the non-gradient terms of the
effective action\cite{boysinlee,frw,avanvega}).

In the absence of an explicit symmetry breaking term, the Goldstone theorem
requires the existence of massless pions, ${\cal{M}}_{\pi}=0$ whenever
${\cal{V}}^1$ for $\phi \neq 0$.
However, this is
{\em not}
the case within our approximation
scheme as it stands.

This situation
can be easily remedied, however,
by noting that the Hartree approximation
becomes exact in the large
$N$
limit. In this limit,
 $\langle \vec{\pi}^2 \rangle
\approx {\cal{O}}(N), \ \langle \chi^2 \rangle \approx
 {\cal{O}}(1), \ \phi^2
\approx {\cal{O}}(N)$. Thus we will approximate further by neglecting the
${\cal{O}}(1/N)$ terms in the formal large
$N$
limit. This further truncation
ensures that the Ward identities are satisfied. We now obtain
\ba
{\cal{V}}^1(t)       & = & \phi(t) \left[m^2(t)+4\lambda \phi^2(t)+
4\lambda \langle \vec{\pi}^2 \rangle(t) \right] -h \label{nu1} \\
{\cal{M}}^2_{\pi}(t) & = & m^2(t)+4\lambda \phi^2(t)+4\lambda
\langle \vec{\pi}^2 \rangle (t) \label{pionmass} \\
{\cal{M}}^2_{\chi}(t)& = & m^2(t)+12\lambda \phi^2(t)+4\lambda
\langle \vec{\pi}^2 \rangle (t) . \label{chimass}
\ea
The Hamiltonian is now quadratic with time dependent self-consistent
masses and Goldstone's Ward identities are satisfied.

Since the evolution Hamiltonian is quadratic in this approximation, we
propose in the Hartree approximation a gaussian density matrix in terms
of the Hartree-Fock states. As a consequence of our assumption of no
cross correlation between $\chi$ and $\vec{\pi}$ , the density matrix
factorizes as
\be
\rho(t) = \rho_{\chi}(t)\otimes \rho_{\pi}(t) \nonumber
\ee

In the
Schr\"{o}edinger
representation the density matrix is most easily
written down by
making use of spatial translational invariance to
decompose the fluctuation fields $\chi(\x,t)$ and $\vp(\x,t)$ into spatial
Fourier modes:

\ba
\chi(\x,t) & = & \frac{1}{\sqrt{\Omega}} \sum_{\k} \chi_{\k}(t)
\exp(-i\k \cdot \x) \\
\vp(\x)    & = & \frac{1}{\sqrt{\Omega}} \sum_{\k} \vp_{\k}
\exp(-i\k \cdot \x) ,
\ea
where we recall that $\chi(\x,t)=\sigma(\x)-\phi(t)$.
We can now use these Fourier modes as the basis in which to write the density
matrices for the sigma and the pions. We will use the following Gaussian
ansatze:

\ba
\rho_{\chi}[\chi,\tilde{\chi},t] & = & \prod_{\vec{k}}
{\cal{N}}_{\chi,k}(t)
\exp\left\{- \left[\frac{A_{\chi,k}(t)}{2\hbar}\chi_k(t)\chi_{-k}(t)+
\frac{A^*_{\chi,k}(t)}{2\hbar}\tilde{\chi}_k(t)\tilde{\chi}_{-k}(t)+
\right. \right.
\nonumber \\
                                 &   & \left. \left.
 \frac{B_{\chi,k}(t)}{\hbar}\chi_k(t)\tilde{\chi}_{-k}(t)
\right]
+\frac{i}{\hbar}p_{\chi,_k}(t)\left(\chi_{-k}(t)-
\tilde{\chi}_{-k}(t)\right) \right\}, \label{chidensitymatrix}
\ea

\ba
\rho_{\vp}[\vp,\tilde{\vp},t] & = & \prod_{\vec{k}} {\cal{N}}_{\pi,k}(t)
\exp\left\{- \left[\frac{A_{\pi,k}(t)}{2\hbar}\vp_k\cdot \vp_{-k}+
\frac{A^*_{\pi,k}(t)}{2\hbar}\tilde{\vp}_k\cdot \tilde{\vp}_{-k}+
\right. \right.
\nonumber \\
                              &   & \left. \left.
\frac{B_{\pi,k}(t)}{\hbar}\vp_k\cdot
\tilde{\vp}_{-k}\right]\right\} . \label{pidensitymatrix}
\ea
The assumption of isospin invariance implies that the kernels
 $A_{\pi,k},\ B_{\pi,k}$
transform as isospin singlets, since these kernels give the two point
 correlation functions. Furthermore, hermiticity of the density matrix
requires that the mixing kernel $B$ be real. The lack of a linear
term in the pion density matrix will become clear below.

The Liouville equation is most conveniently solved in the
Schr\"{o}edinger
representation, in which

\[\Pi_{\chi}(\x)= -i\hbar\frac{\delta}
{\delta \chi} \; \; ; \; \; \Pi^j_{\pi}(\x)= -i\hbar\frac{\delta}
{\delta \pi_j} , \]
\be
i \hbar \frac{\partial \rho(t)}{\partial t} =
\left(H[\Pi_{\chi},\vec{\Pi}_{\vp}, \chi,\vp;t]- H[\tilde{\Pi}_{\chi},
\tilde{\vec{\Pi}}_{\vp}, \tilde{\chi},\tilde{\vp};t]\right)\rho(t) .
\ee

Comparing the terms quadratic, linear and independent of the
fields ($\chi \; ; \; \vp$),  we obtain the following set of differential
equations for the coefficients and the expectation value:

\ba
i\frac{{\dot{\cal{N}}}_{\chi,k}}{{\cal{N}}_{\chi,k}} & = &
\frac{1}{2}(A_{\chi,k}-A^*_{\chi,k}) \label{normchi} \\
i\dot{A}_{\chi,k}                                    & = &
\left[A^2_{\chi,k}-B^2_{\chi,k}-\omega^2_{\chi,k}(t) \right]
 \label{achi} \\
i\dot{B}_{\chi,k}                                    & = &
B_{\chi,k}\left(A_{\chi,k}-A^*_{\chi,k}\right) \label{bchi} \\
\omega^2_{\chi,k}(t)                                 & = &
k^2+{\cal{M}}^2_{\chi}(t) \label{omega2chi}
\ea
\be
\ddot\phi+m^2(t)\phi+4\lambda\phi^3+4\lambda\phi \langle \vec{\pi}^2(\x,t)
\rangle -h =  0 .
\label{fieqnofmotion}
\ee

\ba
i\frac{{\dot{\cal{N}}}_{\pi,k}}{{\cal{N}}_{\pi,k}}  & = &
\frac{1}{2}(A_{\chi,k}-A^*_{\chi,k}) \label{normpi} \\
i\dot{A}_{\pi,k}                                    & = &
\left[A^2_{\pi,k}-B^2_{\pi,k}-\omega^2_{\pi,k}(t) \right]
 \label{api} \\
i\dot{B}_{\pi,k}                                    & = &
B_{\pi,k}\left(A_{\pi,k}-A^*_{\pi,k}\right) \label{bpi} \\
\omega^2_{\pi,k}(t)                                 & = &
k^2+{\cal{M}}^2_{\pi}(t). \label{omega2pi}
\ea

The lack of a linear term in (\ref{pidensitymatrix}) is a consequence
of a lack of a linear term in $\vp$ in the Hartree Hamiltonian, as the
symmetry has been specified to be broken along the sigma direction.

To completely solve for the time evolution, we must specify the initial
conditions. We will {\it assume} that at an initial time
($t=0$) the
system is in {\it local thermodynamic equilibrium} at an initial
temperature
$T$, which we take to be higher than the critical
temperature,$T_c \approx \mbox{ 200 MeV }$,
where we use the
phenomenological couplings and masses to obtain $T_c$.

This assumption thus describes the situation in a
high energy collision in which the central rapidity region is at a
temperature larger than critical,
and thus in the symmetric phase, and such that the
phase transition occurs via the rapid cooling that occurs when the
region in the high temperature phase expands along the beam axis.

The assumption of local thermodynamic equilibrium for the Hartree-Fock
states determines the initial values of the kernels and the expectation
value of the sigma field and its canonical momentum:
\ba
A_{\chi,k}(t=0) & = & \omega_{\chi,k}(0)\coth[\beta \hbar
 \omega_{\chi,k}(0)] \label{iniachi} \\
B_{\chi,k}(t=0) & = & -\frac{\omega_{\chi,k}(0)}{\sinh[\beta \hbar
 \omega_{\chi,k}(0)]} \label{inibchi} \\
A_{\pi,k}(t=0)  & = & \omega_{\pi,k}(0)\coth[\beta \hbar
 \omega_{\pi,k}(0)] \label{iniapi} \\
B_{\pi,k}(t=0)  & = & -\frac{\omega_{\pi,k}(0)}{\sinh[\beta \hbar
 \omega_{\pi,k}(0)]} \label{inibpi} \\
\phi(t=0)       & = & \phi_0 \; \; ; \; \;
\dot{\phi}(t=0)  =  0 , \label{initialfidot}
\ea
with $\beta = 1/k_B T$. We have (arbitrarily) assumed that the expectation
value of the canonical momentum conjugate to the sigma field is zero in
the initial equilibrium ensemble. These initial conditions dictate the
following ansatze for the real and imaginary parts
of the kernels $A_{\pi,k}(t),\ A_{\chi,k}(t)$ in terms of
complex functions ${\cal{A}}_{\pi,k}(t)={\cal{A}}_{R;\pi,k}(t)+
i{\cal{A}}_{I;\pi,k}(t)$ and  ${\cal{A}}_{\chi,k}(t)={\cal{A}}_{R;\chi,k}(t)+
i{\cal{A}}_{I;\chi,k}(t)$\cite{frw}:

\ba
A_{R;\pi,k}(t) & = & {\cal{A}}_{R;\pi,k}(t) \coth[\beta \hbar
 \omega_{\pi,k}(0)]  \label{aresol} \\
B_{\pi,k}(t)   & = & -\frac{{\cal{A}}_{R;\pi,k}(t)}{\sinh[\beta \hbar
 \omega_{\pi,k}(0)]}  \label{bsol} \\
A_{I;\pi,k}(t) & = &  {\cal{A}}_{I;\pi,k}(t) \label{aimsol}
\ea
The differential equation for the complex
function ${\cal{A}}$
can be cast in a more familiar form by a change of variables

\be
{\cal{A}}_{\pi,k}(t) = -i \frac{\dot{\Psi}_{\pi,k}(t)}{\Psi_{\pi,k}(t)} ,
\label{psi}
\ee
with $\Psi_{\pi,k}$ obeying the following Schr\"{o}dinger-like differential
equation, and boundary conditions
\ba
\left[\frac{d^2}{dt^2}+\omega^2_{\pi,k}(t) \right]\Psi_{\pi,k}(t)
                         & = & 0 \label{diffeqnpsi} \\
\Psi_{\pi,k}(t=0)        & = & \frac{1}{\sqrt{\omega_{\pi,k}(0)}}
\; \; ; \; \;
\dot{\Psi}_{\pi,k}(t=0)
                          =  i\sqrt{\omega_{\pi,k}(0)} .
\label{psidot0}
\ea

Since in this approximation the dynamics for the pions and sigma fields
decouple, we will only concentrate on the solution for the pion fields;
the effective time dependent frequencies for the sigma fields are
completely determined by the evolution of the pion correlation functions.
In terms of these functions
we finally find
\be
\langle {\vp}_{k}(t) \cdot {\vp}_{-k}(t) \rangle  =
\frac{N\hbar}{2} \mid \Psi_{\pi,k}(t)\mid^2 \coth\left[
\frac{\hbar \omega_{\pi,k}(0)}{2k_BT} \right].
\label{pioncorrfunc}
\ee
In terms of this two-point correlation function and recognizing that
the $\Psi_{\pi,k}(t)$ only depends on $k^2$, we obtain the following
important correlations :

\ba
\langle \vp^2(\x,t) \rangle       & = & \frac{N\hbar}{4\pi^2}\int dk
k^2 \mid \Psi_{\pi,k}(t)\mid^2 \coth\left[
\frac{\hbar \omega_{\pi,k}(0)}{2k_BT} \right] \label{pi2corrfunc} \\
\langle \vp(\x,t) \cdot \vp(\vec{0},t) \rangle
                                  & = & \frac{N\hbar}{4\pi^2}\int dk
k\frac{\sin(kx)}{x} \mid \Psi_{\pi,k}(t)\mid^2 \coth\left[
\frac{\hbar \omega_{\pi,k}(0)}{2k_BT} \right] .
\label{pispatialcorrfunc} \\
\ea
The presence of the temperature dependent function in the above
expressions
encodes
the finite temperature correlations of the initial
state.
The set of equations (\ref{fieqnofmotion},\ref{diffeqnpsi}) with the
above boundary conditions, completely determine the non-equilibrium
 dynamics in the
Hartree-Fock approximation. We will provide a numerical analysis of
these equations in the next section.

\subsection{\bf Pion production}

In the Hartree approximation, the Hamiltonian is quadratic, and the
fields can be expanded in terms of creation and annihilation of
Hartree-Fock states
\be
{\vp}_{k}(t)  =  \frac{1}{\sqrt{2}} \left(\vec{a}_k
 \Psi^{\dagger}_{\pi,k}(t)+
\vec{a}^{\dagger}_{-k} \Psi_{\pi,k}(t) \right) .
\label{heisfield}
\ee
The creation $\vec{a}^{\dagger}_k$ and annihilation $\vec{a}_k$
operators are {\it independent of time} in the Heisenberg picture and
the mode functions $\Psi_{\pi,k}(t)$ are the solutions to the Hartree
equations (\ref{diffeqnpsi}) which are the Heisenberg equations of motion
in this approximation. The boundary conditions (\ref{psidot0})
correspond to positive frequency particles for $t \leq 0$.
The creation and annihilation operators may be
written in terms of the Heisenberg fields (\ref{heisfield}) and their
canonical momenta. Passing on to the
Schr\"{o}edinger
picture at
time $t=0$, we can relate the
Schr\"{o}edinger
picture operators at time
$t$ to those at time $t=0$ via a Bogoliubov transformation:
\be
\vec{a}_k (t)  =   {\cal{F}}_{+,k}(t) \vec{a}_k(0) +
{\cal{F}}_{-,k}(t)
\vec{a}^{\dagger}_{-k} (0) \label{bogoltrans}
\ee
with

\ba
\mid {\cal{F}}_{+,k}(t) \mid^2    & = &
\frac{1}{4}\frac{\mid \Psi_{\pi,k}(t) \mid^2}
{\mid \Psi_{\pi,k}(0) \mid^2}
 \left[1+
\frac{\mid {\dot{\Psi}}_{\pi,k}(t)\mid^2}{\omega^2_{\pi,k}(0)
\mid \Psi_{\pi,k}(t) \mid^2}\right] + \frac{1}{2}
 \label{bogol} \\
\mid {\cal{F}}_{+,k}(t) \mid^2    & - &
\mid {\cal{F}}_{-,k}(t) \mid^2 =1 .
\nonumber
\ea

At any time $t$ the expectation value of the number operator for pions
(in each
$k$-mode) is
\be
\langle N_{\pi,k}(t) \rangle = \frac{Tr \vec{a}^{\dagger}_{\pi,k}(t)\cdot
 \vec{a}_{\pi,k}(t) \rho(0)}{Tr\rho(0)} =
\frac{Tr \vec{a}^{\dagger}_{\pi,k}(0)\cdot
 \vec{a}_{\pi,k}(0) \rho(t)}{Tr\rho(t)} .
\label{pionnumber}
\ee
After some straightforward algebra we find
\be
\langle N_{\pi,k} \rangle(t) = (2
\mid{\cal{F}}_{+,k}(t,t_o)\mid^2 -1)\langle N_{\pi,k}\rangle(0)+
\left(\mid{\cal{F}}_{+,k}(t,t_o)\mid^2 -1\right) .
\label{numboft}
\ee
The first term represents the ``induced'' and the second term the
``spontaneous'' particle production. In this approximation, particle
production is a consequence of parametric amplification. The
Hartree-Fock states are examples of squeezed states, and the density matrix
is a ``squeezed'' density matrix. The squeeze parameter (the
ratio of the kernels at a time $t$
to those at time $t=0$)  is time
dependent and determines the time evolution of the states and
density matrix. The relation between squeezed states and pion production
has been advanced by Kogan\cite{kogan} although not in the context of an
initial thermal density matrix.

Thus far we have established the formalism to study the non-equilibrium
evolution during the phase transition.
A question of interpretation must be clarified before proceeding further.
Our description, in terms of a statistical density matrix, describes an
isospin invariant mixed state, and thus does not prefer one isospin
direction over another. A real experiment will furnish one realization
of all the available states mixed in the density matrix. However, if
the pion correlation functions become long ranged (as a statistical
average) it is clear that in a particular realization, at least one
isospin component is becoming correlated over large distances, thus it is
in this statistical sense that our results should be understood.

This concludes our discussion of the formalism we will use to study the
non-equilibrium evolution of the pion system. We now turn to a numerical
analysis of the problem.

\section{\bf Numerical Analysis}

The phenomenological set of parameters that define the linear
sigma model as an effective low energy theory are as follows
(we will be somewhat cavalier about the precise value of these
parameters as we are interested in the more
robust features of the
pion correlations)

\ba
 M_{\sigma} \approx 600 \mbox{ MeV }     & & \; \; ; \; \;
  f_{\pi} \approx
95 \mbox{ MeV } \; \; ; \; \; \lambda \approx 4.5 \nonumber \\
h \approx (120 \mbox{ MeV })^3             & & \; \; ; \; \;
 T_c \approx 200 \mbox{ MeV } .
\label{parameters}
\ea

The above value of the critical temperature differs somewhat from the
lattice estimates ($T_c \approx 150 \mbox{ MeV }$), and
 our definition of
$\lambda$ differs by a factor four from that given
elsewhere\cite{wilraj2,pisarski}.

The first thing to notice
is that this is a {\it strongly} coupled theory, and unlike our
previous studies of the dynamics of
phase transitions\cite{boysinlee,danhec} we expect
the relevant time scales to be much {\it shorter} than
in weakly coupled theories.

We must also notice that the linear sigma model is an effective low-energy {\it
cutoff} theory. There are two physically important factors that limit the value
of the cutoff: i) this effective theory neglects the influence of the nucleons,
and in the Hartree approximation, the vector resonances are missed. These two
features imply a cut-off of the order of about 2 GeV; ii) the second issue is
that of the triviality bound. Assuming that the value of the coupling is
determined at energies of the order of $M_{\sigma}$, its very large value
implies that the cutoff should not be much larger than about 4-5 GeV,
since otherwise
the theory will be dangerously close to the Landau pole. From the technical
standpoint this is a more important issue since in order to write the
renormalized equations of motion we need the  ratio between bare and
renormalized couplings. In the Hartree approximation this is the ``wave
function renormalization constant'' for the composite operator ${\vp}^2$.

Thus we use a cutoff $\Lambda=2 \mbox{ GeV }$. The issue of the cutoff
is an important one since $\langle {\vp}^2 \rangle$ requires renormalization,
and in principle we should write down renormalized equations of motion.
In the limit when the cutoff is taken to infinity the resulting evolution
should be insensitive to the cutoff. However the chosen cutoff is not
very much larger than other scales in the problem and the ``renormalized''
equations will yield solutions that are cutoff sensitive. However, this
sensitivity will manifest itself on distance scales of the order of $0.1$ fm
or smaller, and we are interested in detecting correlations over many
fermis. The {\it size} of the correlated regions and the time scales for
their growth will be determined by
the long wavelength unstable modes\cite{boysinlee} (see below), and
thus should be fairly insensitive to the momentum scales
near the cutoff. The short distance features of the correlation functions,
such as {\it e.g.} the {\it amplitude} of the fluctuations will, however, be
rather sensitive to the cutoff.

The most severe ultraviolet divergence in the composite operator
${\vp}^2$
is proportional to $\Lambda^2$.
This divergence
is usually handled by a subtraction. We will subtract this term
(including the temperature factors) in a renormalization of the mass
at $t=0$.
Thus

\be
m^2_B(t=0)+4\lambda\langle \vp^2 \rangle (t=0;T) =m^2_R(t=0;T)
\label{renormass}
\ee
where we made explicit the temperature dependence of ${\vp}^2$ and
$m^2_R(t=0;T)$.
Furthermore we will parametrize the time dependent mass term as
\be
m^2_R(t)=\frac{M_{\sigma}^2}{2}\left[\frac{T^2}{T^2_c}
\exp{\left[-2\frac{t}{t_r}\right]}-1\right] \Theta(t)+
\frac{M_{\sigma}^2}{2}\left[\frac{T^2}{T^2_c}
-1\right] \Theta(-t) .  \label{massoft}
\ee

This parametrization incorporates the dynamics of the expansion and cooling
processes in the plasma in a phenomenological way. It allows for the system
to cool down with an effective temperature given by:

\be
T_{eff}(t) = T\exp{\left[-\frac{t}{t_r}\right]} \label{tempoft}
\ee
where $T$ is the initial value of the temperature in the central rapidity
region, and $t_{r}$ is a relaxation time. This parametrization also allows us
to study a ``quench'' corresponding to the limiting case $t_r=0$.

It is convenient to introduce the natural scale
 $\mbox{ fm }^{-1} \approx 200 \mbox{ MeV } = M_F$ and define the following
dimensionless variables
\ba
\phi(t) & = &  M_F f(t) \; \; ; \; \; \Psi_{\pi,k}(t) =
\frac{\psi_q(\tau)}{\sqrt{M_F}} \nonumber \\
 k      & = & M_F q \; \; ; \; \; t =
 \frac{\tau}{M_F} \; \; ; \; \; t_r =
 \frac{\tau_r}{M_F} \; \; ; \; \; x= \frac{z}{M_F} .
\ea

In these units
\ba
\frac{\Lambda}{M_F} & = & 10 \; \; ; \; \; \frac{M_{\sigma}}{M_F} =3
\; \; ; \; \; H=\frac{h}{M_F^3} \approx 0.22
\label{numbers} \\
\frac{\omega^2_{\pi,k}(0)}{M_F^2}
                    & = & W^2_q =
 q^2+\frac{9}{2}\left[\frac{T^2}{T_c^2}
-1\right]+4 \lambda f^2(0) \label{omegadim}
\ea

Thus we have to solve simultaneously the Hartree set of equations:
\ba
 \frac{d^2 f}{d\tau^2}+\frac{9}{2}f\left[\frac{T^2}{T_c^2}
\exp{\left[-2\frac{\tau}{\tau_r}\right]}
-1\right]+4 \lambda f^3 + 4f \lambda \Sigma(0,\tau)-H
& = & 0 \label{dimfieqn} \\
\left\{ \frac{d^2}{d\tau^2}+q^2+\frac{9}{2}\left[\frac{T^2}{T_c^2}
\exp{\left[-2\frac{\tau}{\tau_r}\right]}
-1\right]+4 \lambda f^2(\tau) + 4 \lambda \Sigma(0,\tau) \right\}
 \psi_q(\tau)
& = & 0 \label{dimpsieqn}
\ea
\ba
\Sigma(z,\tau) & = & \langle \vp(\x,t) \cdot \vp(\vec{0},t) \rangle /
 M_F^2 \nonumber \\
               & = & \frac{3}{4\pi^2} \int^{10}_{0}dq q
\frac{\sin(qz)}{z}  (\mid \psi_q(\tau) \mid^2-
\mid \psi_q(0) \mid^2) \coth\left[\frac{W_q}{10 T(\mbox{ GeV })}\right]
\label{Sigma}
\ea
with the boundary conditions
\ba
f(0) & = & f_0 \; \; ; \; \; \frac{df(0)}{d\tau} = 0 \label{fibound} \\
\psi_q(0)
     & = & \frac{1}{\sqrt{W_q}} \; \; ; \; \; \frac{d \psi_q}{d\tau}(0) =
i\sqrt{W_q} \label{psibound}
\ea
Finally, once we find the Hartree mode functions, we can compute the
total pion density as a function of time:
\be
\frac{N_{\pi}(t)}{\Omega} =
\frac{\overline{N}_{\pi}(\tau)}{(\mbox{ fm })^3}
 = \frac{1}{2\pi^2 \mbox{ fm }^3} \int^{10}_{0} dq~
 q^2 \langle N_{\pi,q}(\tau) \rangle \label{piondensity}
\ee
with $\langle N_{\pi,q}(\tau) \rangle $ given by (\ref{numboft}) in terms
of the dimensionless variables.

The mechanism of domain formation and growth is the  fast time evolution
of the unstable modes\cite{boysinlee}.

 Let us consider first the case of
$H=0$. Then $f=0$
is a fixed point of the evolution equation for $f$ and
corresponds to cooling down from the symmetric (disordered) phase in the
absence of explicit symmetry breaking perturbations. Let us consider
the simpler situation of a quench ($\tau_r=0$). The  equation for the
Hartree mode functions (\ref{dimpsieqn}) shows that for $q^2 < 9/2$ the
corresponding modes are unstable at early times and grow exponentially.

This growth feeds back on $\Sigma(\tau)$ which begins to grow and tends
to overcome the instability. As the unstable fluctuations grow,
only longer
wavelengths remain unstable, until the time when
$4 \lambda \Sigma(\tau) \approx 9/2$,
at which point no wavelength is
unstable.
The modes will continue to grow however,
because the derivatives will be
fairly large, but since the instabilities
will be overcome beyond this time the modes will have an oscillatory
 behavior.

We expect then that the fluctuations will grow during the time for which
there are instabilities.
This time scale depends on the value of the
coupling;
 for very small coupling, $\Sigma(\tau)$ will have to grow for
a long time before $4 \lambda \Sigma(\tau) \approx 9/2$ and
the instabilities are shut-off. On the other hand, for strong coupling
this time scale will be rather small, and domains will not have much time
to grow.

It is clear that allowing for a non-zero magnetic field or $f(0) \neq 0$
will help to shut off the instabilities at earlier times, thus making
the domains even smaller.

For typical relaxation times ($\approx 1-2 \rm{\mbox{ fm/c }}$),
 domains will not grow too large either because the fast growth of
the fluctuations will  catch up with the relaxing modes and
 shut-off the instabilities
(when $T_{eff}(\tau) < T_c$) on short time scales.
Thus we expect that for a non-zero  relaxation time
 $\tau_r \approx \rm{\mbox{ fm/c }}$, domains will not grow too
 large either because
the fluctuations will shut-off the instabilities
(when $T_{eff}(\tau) < T_c$) on shorter time scales (this argument will
be confirmed numerically shortly).

We conclude from this analysis that the optimal situation for which large DCC
regions can grow
corresponds to a quench from the symmetric phase in the absence of a
magnetic field.

Figure (1.a) shows $4 \lambda \Sigma(\tau) \mbox{ vs } \tau$,
figure (1.b) shows $\Sigma(z,\tau=1;3;5) \mbox{ vs } z$ and
and figure (1.c) shows $\overline{N}_{\pi}(\tau) \mbox{ vs } \tau$ for
 the
following values of the parameters: $T_c = 200 \mbox{ MeV } \; \; ; \; \;
T = 250 \mbox{ MeV }\; \; ; \; \; \tau_r=0 \; \; ; \; \; f(0)=0 \; \; ;
\; \; H=0 \; \; ; \; \;  \lambda=4.5 $ corresponding to a quench from the
symmetric phase at a temperature slightly above the critical temperature
 and no  magnetic field.

 We clearly see that the fluctuations grow to
overcome the instability in times $\approx 1 \mbox{ fm }$ and the
domains never get bigger than about $\approx 1.5 \mbox{ fm }$.
Figure (1.c) shows that the number of pions per cubic fermi is about
0.15 at the initial time (equilibrium value) and grows to about
0.2 in times about 1-2 fermis after the quench. This pion density is
thus consistent with having only a few pions in a pion-size correlation
volume.

 Figures (2.a-c) show again the same functions but now we let the system
relax from an initial temperature $T=T_c$ with a relaxation time
 $t_r=1 \mbox{ fm }/c$. Notice that now the fluctuations grow less
rapidly as they are modulated by the relaxation time, but again on a time
scale of the order of a fm/c, they become big enough to shut-off the
long wavelength instabilities. Figure (2.b) shows
 $\Sigma(z,\tau=1;3;5;7) \mbox{ vs } z$.
 Once again, pions are correlated over
distances of the order of a Fermi.
The reason the fluctuations grow so quickly and thus shut off the growth of the
unstable modes so quickly is the strongly coupled nature of the theory.

The possibility of long range correlations exists if the initial
state is in {\it equilibrium at the critical temperature}. In this
situation there are already  long range correlations in the initial
state that will remain for some time as the temperature factors enhance
the contributions for long wavelength modes since the Boltzmann factor
$\approx 1/k$ for long wavelength fluctuations. Figures (3.a-c) show
this situation for the values of the parameters  $ T=T_c = 200
\mbox{ MeV } \; \; ; \; \;
 \tau_r=0 \; \; ; \; \; f(0)=0 \; \; ;
\; \; H=0 \; \; ; \; \;  \lambda=4.5 $. Figure (3.b) shows
 $\Sigma(z,\tau=1;2) \mbox{ vs } z$. In this case the number of
pions per cubic fermi in the initial state is $\approx 0.12$ and
reaches a maximum of about $0.17$ within times of the order of a fermi/c.
 The pions, however, are correlated
over distances of about $4-5 \mbox { fm }$ with a large number of
pions per correlation volume $\approx 50$. These large correlation
volumes are a consequence of the initial long range correlations.
This is the situation proposed by Gavin, Gocksch and Pisarski for the
possibility of formation of large domains, as there is a ``massless''
particle in the initial state.

We believe that this situation is not very likely as the central
rapidity region must remain in equilibrium at (or very close) to the
critical temperature before the quench occurs.

To contrast this situation with that of a weakly coupled theory, figures
(4.a-c) show the same functions with the following values of the parameters
 $ \lambda=10^{-6} \; \; ; \; \; T_c = 200 \mbox{ MeV } \; \; ; \; \;
T = 250 \mbox{ MeV }\; \; ; \; \; \tau_r=0 \; \; ; \; \; f(0)=0 \; \; ;
\; \; H=0$. Now the fluctuations are negligible up to times of about 4
fm/c, during which time the correlation functions grow (figure 4.b)
and pions become correlated over distances of the order 4-5 Fermis. As
shown in figure (4.c) the number of pions per cubic fermi becomes enormous,
a consequence of a large parametric amplification. In this extremely
weakly coupled theory, the situation of a quench from above the critical
temperature to almost zero temperature does produce a large number of
coherent pions and domains which are much larger than typical pion sizes.
This is precisely the situation studied previously\cite{boysinlee}
 within a different context.

We have analyzed numerically many different situations in the strongly
coupled case ($\lambda \approx 4-5$) including the magnetic field and
letting the expectation value of the sigma field ``roll-down'' etc, and
in {\it all} of these cases in which the initial temperature is higher
than the critical (between $ 10-20 \%$ higher) we find the
 common feature that the time
 and spatial scales of correlations are  $\approx 1 \mbox{ fm }$.  Thus
it seems that within this approach the strongly coupled linear sigma
model is incapable of generating large domains of correlated pions.

\section{\bf Discussions and Conclusions:}

Our study differs in many qualitative and quantitave ways from previous
studies. In particular we incorporate both quantum and thermal
fluctuations and correlations in the initial state. In previous studies
it was argued that because one is interested in long-wavelength
fluctuations these may be taken as {\it classical} and the classical
evolution equations (with correlations functions replaced by spatial
averages) were studied. We think that it is important to quantify why
and when the long-wavelength fluctuations are classical within the
present approximation scheme.
This may be
seen from the temperature factors in the Hartree propagators. These
are typically (incorporating now the appropriate powers of $\hbar$):
\[\hbar \coth\left[\frac{\hbar \omega_k}{2k_BT}\right] . \]
Thus, modes with wavelength $k$ and energies $\omega_k$ are classical
when
$\hbar \omega_k \ll k_BT$  and yield a contribution to the propagator

\[\hbar \coth\left[\frac{\hbar \omega_k}{2k_BT}\right]
 \approx 2k_B T/ \omega_k\]

(notice the cancellation of the $\hbar$). For long-wavelength components
this happens when

\[ \frac{M_{\sigma}^2}{T^2}[T^2/T^2_c-1] \ll 1 , \]
because the ``thermal mass'' (squared) for the excitations in the
 heat bath is
$\frac{M_{\sigma}^2}{2}[T^2/T^2_c-1]$.
  For the phenomenological values of
$M_{\sigma}$ and $T_c$, the ``classical'' limit is obtained when

\be
[\frac{T^2}{T_c^2}-1] \ll 0.1,
\ee
that is, when the initial state is in {\it equilibrium} at a temperature
that is {\em extremely}
close to the critical temperature. This is the situation that is
shown in figures (3.a-c) where, indeed, we obtain very large correlated
domains that were already present in the initial state after a quench
from the critical temperature all the way to zero temperature.

After an energetic collision it seems rather unlikely that the central
region will be so close to the critical temperature.
If the temperature is higher than critical, in order for the
system to cool down to the critical temperature (or very near to it)
and to remain in {\it local thermodynamic equilibrium},
very long relaxation times are
needed, as the long-wavelength modes are typically critically slowed
down during a transition. Long relaxation times will allow the
fluctuations to shut off the instabilities as they begin to grow and the
system will lose its long range correlations.
This was the original
argument that discarded an equilibrium situation as a candidate for
large domains. Furthermore,
 typical heavy ion collisions or high energy processes will
not allow long relaxation times (typically of a few fermis/c). Thus
we believe that in most generic situations, a classical approximation
for the long wavelength modes is not reliable {\it in the Hartree}
approximation.
We should make here a very important point. We are {\it not} saying that
large coherent fluctuations cannot be treated semiclassically. They can.
What
we are asserting with the above analysis is that
within the Hartree approximation, long wavelength excitations cannot
be treated as classical. The Hartree approximation in the form used by
these (and most other) authors {\it does not} capture correctly the
physics of coherent semiclassical non-perturbative configurations.

Thus although the most promising situation, within the model under
investigation, is a quench from the critical temperature (or very close
to it) down to zero temperature, it is our impression that this scenario
is physically highly unlikely.

There is another very tantalizing possibility for the formation of
large correlated pion domains within the {\it linear} sigma model
and
that is via the creation of a
critical droplet that will complete the phase transition (first order
in this case) via the process of thermal activation over a ``free
energy'' barrier.
The small magnetic field (resulting from the small up and down quark
masses) introduces a small metastability\cite{note}.
 The classical equations of
motion allow for a solution in which the pion field is zero everywhere
and a droplet in the sigma field (this is the O(3) symmetric bounce
responsible of thermal activation in scalar metastable theories in
three dimensions\cite{linde}). Using a spherically symmetric
ansatz
for a sigma droplet of radius R
\be
\sigma_{cl}(r) \approx f_{\pi} \tanh[M_{\sigma}(r-R)],
\ee
and assuming, for the sake of argument,
that the thin-wall approximation
is reliable, we obtain an approximate form for the energy of the
droplet
\be
E \approx 4\pi R^2 f^2_{\pi}M_{\sigma}-\frac{4\pi}{3}R^3
h f_{\pi}.
\ee
The
critical radius is thus (this approximation is clearly
reliable only as  an  order-of-magnitude estimate)
\[ R_c \approx  3-5 \rm{\mbox{ fm }}. \]
By considering the fluctuations of the pions around this configuration,
it is conceivable (although we cannot provide a more convincing argument
at this stage) that the unstable mode of the droplet (dilation) that makes
the droplet grow to complete the phase transition via thermal activation,
produces a large amount of correlated pions. This scenario, however,
requires supercooling (the false vacuum to be trapped) which again
requires long relaxation times (again unlikely for strong coupling).

As argued above, this possibility cannot be studied via a
 Hartree approximation which
only provides a (select) resummation of  the perturbative expansion and
is probably reliable only for short times, before non-perturbative
configurations truly develop.

Thus we conclude that although our analysis provides a negative answer
to the question of the possibility of large correlated domains near the
chiral phase transition, these results are
valid only within the Hartree approximation
of the linear sigma model. There are several conceivable
possibilities
that would have to be studied thoroughly
before any conclusions are reached: i) perhaps the linear sigma model
is not a good candidate for studying the {\it dynamics} of the chiral
phase transition (although it describes the universality class for the
static properties) ii) there are large coherent field configurations
(droplets) that are not captured in the Hartree approximation. This
possibility is rather likely and is closer to the scenario envisaged by
Bjorken, Kowalski and Taylor\cite{dcc5}. These semiclassical coherent
configurations may be responsible for large
regions of correlated pions. An important ingredient in this latter
case must be a deeper understanding of the dynamical (relaxation)
time scales, for which a deeper understanding of the underlying strongly
interacting theory is needed. A particularly relevant question is whether
such a strongly coupled theory can yield enough supercooling so as to
produce such a configuration.

We believe that these two possibilities must be studied further to
give an unequivocal answer to the question of large correlated domains.
We are currently studying the second possibility in more detail.

\centerline{\bf Acknowledgements}

D. B. was supported through N.S.F. grant: PHY-9302534. This work was also
partially supported by a France-U.S.A.  binational collaboration between
 CNRS
and N.S.F. through N.S.F. grant No: INT-9216755. R.H. was supported in part by
D.O.E. contract $\#$ DOE-ER/408682. The authors acknowledge a Grant from
the Pittsburgh Supercomputer Center No: PHY930049P. D.B. would like to
thank R. Willey, L. McLerran, K. Kowalski, and R. Pisarski
 for illuminating conversations and suggestions and M. Madrid
for computational assistance.
R.H. would like to thank Charles Thorn for some
useful discussions.

\newpage

{\bf Figure Captions:}

\underline{\bf Figure 1.a:}

 $4 \lambda \Sigma(\tau) \mbox{ vs } \tau$.
 $T_c = 200 \mbox{ MeV } \; \; ; \; \;
T = 250 \mbox{ MeV }\; \; ; \; \; \tau_r=0 \; \; ; \; \; f(0)=0 \; \; ;
\; \; H=0 \; \; ; \; \;  \lambda=4.5 $

\underline{\bf Figure 1.b:}
 $\Sigma(z,\tau=1;3;5) \mbox{ vs } z$ for the same values of the
parameters as in figure (1.a) larger values of time correspond to larger
amplitudes at the origin.

\underline{\bf Figure 1.c:}
$\overline{N}_{\pi}(\tau) \mbox{ vs } \tau$ for the same values of the
parameters as in figure (1.a). $\overline{N}_{\pi}(0)=0.15$.

\underline{\bf Figure 2.a:}

 $4 \lambda \Sigma(\tau) \mbox{ vs } \tau$.
 $T= 250 \mbox{ MeV } \; \; \;
T_c = 200 \mbox{ MeV }\; \; ; \; \; \tau_r=1
 \; \; ; \; \; f(0)=0 \; \; ;
\; \; H=0 \; \; ; \; \;  \lambda=4.5 $

\underline{\bf Figure 2.b:}
 $\Sigma(z,\tau=1;3;5;7) \mbox{ vs } z$ for the same values of the
parameters as in figure (2.a) larger values of time correspond to larger
amplitudes at the origin.

\underline{\bf Figure 2.c:}
$\overline{N}_{\pi}(\tau) \mbox{ vs } \tau$ for the same values of the
parameters as in figure (2.a). $\overline{N}_{\pi}(0)=0.15$.

\underline{\bf Figure 3.a:}

 $4 \lambda \Sigma(\tau) \mbox{ vs } \tau$.
 $T= T_c = 200 \mbox{ MeV }\; \; ; \; \; \tau_r=0
 \; \; ; \; \; f(0)=0 \; \; ;
\; \; H=0 \; \; ; \; \;  \lambda=4.5 $

\underline{\bf Figure 3.b:}
 $\Sigma(z,\tau=1;2) \mbox{ vs } z$ for the same values of the
parameters as in figure (3.a) larger values of time correspond to larger
amplitudes at the origin.

\underline{\bf Figure 3.c:}
$\overline{N}_{\pi}(\tau) \mbox{ vs } \tau$ for the same values of the
parameters as in figure (3.a). $\overline{N}_{\pi}(0)=0.12$.

\underline{\bf Figure 4.a:}

 $4 \lambda \Sigma(\tau) \mbox{ vs } \tau$.
 $T_c = 200 \mbox{ MeV } \; \; ; \; \;
T = 250 \mbox{ MeV }\; \; ; \; \; \tau_r=0 \; \; ; \; \; f(0)=0 \; \; ;
\; \; H=0 \; \; ; \; \;  \lambda=10^{-6}$

\underline{\bf Figure 4.b:}
 $\Sigma(z,\tau=4;6) \mbox{ vs } z$ for the same values of the
parameters as in figure (4.a) larger values of time correspond to larger
amplitudes at the origin.

\underline{\bf Figure 4.c:}
$\overline{N}_{\pi}(\tau) \mbox{ vs } \tau$ for the same values of the
parameters as in figure (4.a). $\overline{N}_{\pi}(0)=0.15$.


\begin{thebibliography}{99}
\bibitem{note} Gavin and M\"{u}ller\cite{gavin} have recently reported
on the possibility that droplets of the metastable phase may be relevant.
\end{thebibliography}
\end{document}